# Equilibrium velocity distribution of low-energy particles in spherically symmetric gravitational field

Jian-Miin Liu*
Department of Physics, Nanjing University
Nanjing, The People's Republic of China
*On leave, E-mail address: liu@phys.uri.edu

Abstract

In the presence of a spherically symmetric gravitational field, according to Einstein's theory of gravitation, space and time co-join to be a four-dimensional Schwarzschild manifold. We find the velocity space at any point on this manifold. It has an affine structure in the usual (Newtonnian) velocity-coordinate system and the Euclidean structure in the so-called doubly-primed velocity-coordinate system. The transformation between these two velocity-coordinate systems is determinable. Maxwell derived his non-relativistic equilibrium velocity distribution relying on two assumptions, (a) the velocity distribution function is spherically symmetric and (b) the x-, y- and z- components of velocity are statistically independent. Assumptions (a) and (b) reflect structural characteristics of the pre-relativistic velocity space which has the Euclidean structure in the usual velocity-coordinate system. Borrowing Maxwell's derivations in the doubly-primed velocity-coordinate system and using the determined transformation, we obtain the equilibrium velocity distribution of low-energy particles in the spherically symmetric gravitational field. It is no longer the Maxwellian predicted by calculations based on Newton's theory of gravitation and classical statistical mechanics. The obtained velocity distribution is used to produce the rate for low-energy particles in a container with a leak placed in the spherically symmetric gravitational field to escape. The escape rate depends on both position and orientation of the leak. The obtained equilibrium velocity distribution and produced escape rate of low-energy particles in the spherically symmetric gravitational field can serve as the tests of Einstein's theory of gravitation.

1. Introduction

    Gravity is a fundamental interaction of Nature. Differing from all other interactions, it possesses a specific feature that the movement of a body under gravitational interaction is independent of mass of the body. In other words, different bodies (at the same place and moment) in a gravitational field fall with the same acceleration regardless of their masses. This feature has been known for almost four hundred years. Galileo Galilei (1564-1642) was the first evidently to show this feature through his famous Leaning Tower of Pisa free-falling experiment. Later, other physicists achieved more accurate experimental verification for the feature [1-4]. This feature of gravity, in Einstein's hands, was converted to a basic principle of gravitational physics, the equivalence principle [5,6]. It says: the presence of gravitational force must be locally equivalent to the use of non-inertial frame of reference, where the word "locally" means "in a sufficiently small region of space and time". According to the equivalence principle, an observer in an enclosed elevator at rest in a gravitational field may have striking thought that there was no gravitational field but this elevator being accelerated in an upward direction. Postulating that gravity-free space and time in non-inertial frame of reference become curved with respect to those in inertial frame of reference, relying upon the equivalence principle, Einstein developed his Riemann-geometric description of gravity. In his theory of gravitation, space and time co-join to be a four-dimensional space-time; this space-time becomes curved in the presence of gravitational field; the line element of the curved space-time can be obtained by solving the Einstein field equation; any body in this curved space-time will move along a geodesic line if no non-gravitational force acts on it.

    The Einstein field equation consists of ten non-linear partial differential equations joined together. Mathematically, a set of these partial differential equations for a physical situation is always to be solved

under some suitable boundary conditions. We know that, in any inertial frame of reference, the metric of four-dimensional space-time is the Minkowskian (here we leave the question of its possible correction out of consideration) in the absence of gravitational field. That can be taken as the boundary conditions for solving the Einstein field equation in any inertial frame of reference. For any non-inertial frame of reference, we have no such knowledge about the metric structure of its space-time in the absence or the presence of gravitational field that we can take as the boundary conditions for solving the Einstein field equation in it. In this sense, Einstein's theory of gravitation is a theory effective in the inertial frames of reference. Einstein's theory of gravitation is a great development of Newton's one. Besides, it directs us a way to discuss the influences of some gravitational fields on the laws of non-gravitational physics in the inertial frames of reference.

Due to mathematical difficulties, only for a quite few of physical situations we can exactly solve the Einstein field equation. One of these situations is of spherically symmetric gravitational field. In this paper, we are concerned with the influence of a spherically symmetric gravitational field on Maxwell's velocity distribution of low-energy particles, namely the equilibrium velocity distribution of low-energy particles in the spherically symmetric gravitational field. Suppose we have N identical classical particles of low energy in a relatively small volume located at point $(x^1, x^2, x^3)$ in the spherically symmetric gravitational field produced by mass M. Our question is: What is the equilibrium velocity distribution of these N particles? As we know, the answer is the Maxwellian velocity distribution if our calculations are based on Newton's theory of gravitation and classical statistical mechanics. However, the answer will be quite different if we base our calculations on Einstein's theory of gravitation and classical statistical mechanics. The paper contains seven sections: Introduction, The Schwarzschild manifold, Gravitational proper time and gravitational proper distance, The velocity space, Equilibrium velocity distribution of low-energy particles, Leakage of particles in spherically symmetric gravitational field, and Concluding remarks.

2. The Schwarzschild manifold

In inertial frame of reference, the presence of a spherically symmetric gravitational field produced by mass M, according to Einstein's theory of gravitation, makes the four-dimensional space-time be a Schwarzschild manifold of metric tensor [5-9],

$$\begin{pmatrix} 1-r_s/r & 0 & 0 & 0 \\ 0 & -(1-r_s/r)^{-1} & 0 & 0 \\ 0 & 0 & -r^2 & 0 \\ 0 & 0 & 0 & -r^2\sin^2\theta \end{pmatrix}, \qquad (1)$$

in the spherical coordinate system $\{x^0, r, \theta, \phi\}$, $x^0=ct$, with the origin at which mass M is located, where

$$r_s = \frac{2GM}{c^2}, \qquad (2)$$

G is the Newtonian gravitational constant and c is the speed of light. The line element of the Schwarzschild manifold at point $(ct, r, \theta, \phi)$ is

$$ds^2 = (1-\frac{r_s}{r})c^2dt^2 - (1-\frac{r_s}{r})^{-1}dr^2 - r^2d\theta^2 - r^2\sin^2\theta d\phi^2 . \qquad (3)$$

We can represent this Schwarzschild manifold in the Cartesian coordinate system $\{x^0, x^1, x^2, x^3\}$,

$$\begin{aligned} x^1 &= r\sin\theta\cos\phi, \\ x^2 &= r\sin\theta\sin\phi, \\ x^3 &= r\cos\theta . \end{aligned} \qquad (4)$$

Its line element at point (ct, $x^1$, $x^2$, $x^3$) now shapes in

$$ds^2 = g_{ij}(x)dx^i dx^j, \quad i,j=0,1,2,3, \tag{5a}$$

with metric tensor

$$g_{00}(x) = 1 - \frac{r_s}{r}, \tag{5b}$$

$$g_{r0}(x) = g_{0r}(x) = 0, \quad r=1,2,3, \tag{5c}$$

$$g_{rs}(x) = -\delta_{rs} - \frac{r_s/r}{1 - r_s/r}\frac{x^r x^s}{r^2}, \quad r,s=1,2,3, \tag{5d}$$

where $r^2 = (x^1)^2 + (x^2)^2 + (x^3)^2$. More explicitly,

$$ds^2 = (1 - \frac{r_s}{r})c^2 dt^2 + g_{rs}(x)dx^r dx^s, \quad r,s=1,2,3, \tag{6a}$$

$$g_{rs}(x) = -\begin{pmatrix} 1 + \frac{r_s/r}{1 - r_s/r}\frac{x^1 x^1}{r^2} & \frac{r_s/r}{1 - r_s/r}\frac{x^1 x^2}{r^2} & \frac{r_s/r}{1 - r_s/r}\frac{x^1 x^3}{r^2} \\ \frac{r_s/r}{1 - r_s/r}\frac{x^2 x^1}{r^2} & 1 + \frac{r_s/r}{1 - r_s/r}\frac{x^2 x^2}{r^2} & \frac{r_s/r}{1 - r_s/r}\frac{x^2 x^3}{r^2} \\ \frac{r_s/r}{1 - r_s/r}\frac{x^3 x^1}{r^2} & \frac{r_s/r}{1 - r_s/r}\frac{x^3 x^2}{r^2} & 1 + \frac{r_s/r}{1 - r_s/r}\frac{x^3 x^3}{r^2} \end{pmatrix}. \tag{6b}$$

The Schwarzschild manifold is static not varying with time. It approaches the Minkowskian as r goes to infinity. It becomes the Minkowskian everywhere when the spherically symmetric gravitational field is turned off, i.e. M=0.

3. Gravitational proper time and gravitational proper distance

Let us consider an n-dimensional Riemannian manifold,

$$ds^2 = g_{\mu\nu}(x)dx^\mu dx^\nu, \quad \mu,\nu=1,2,\text{------},n,$$

in the coordinate system $\{x^\mu\}$, $\mu=1,2,\text{------},n$, and an arbitrary point $(x^1_0, x^2_0, \text{------}, x^n_0)$ on it. In view of differential geometry, a sufficiently small immediate vicinity at the point on the Riemannian manifold can be treated as a one to lie on the tangent hyper-plane attached to the point, where the point is a contact point of the Riemannian manifold and the tangent hyper-plane and at the point the tangent hyper-plane is centered. The tangent hyper-plane is an affine space with a constant metric tensor, $g_{\mu\nu}(x_0)$ [10]. For this affine space, we can introduce a coordinate system, say $\{x''^\mu\}$, $\mu=1,2,\text{------},n$, called the doubly-primed coordinate system, such that the affine space has the Euclidean line element,

$$ds^2 = \delta_{\mu\nu}dx''^\mu dx''^\nu, \quad \mu,\nu=1,2,\text{------},n,$$

in this coordinate system. So, the sufficiently small immediate vicinity at the point on the Riemannian manifold can be represented in the doubly-primed coordinate system and it has the Euclidean metric structure in this coordinate system. We therefore come to the statement that, for an arbitrary point on a Riemannian manifold, there must exist a locally valid doubly-primed coordinate system in which the point is represented by (0, 0, ------, 0) and the Riemannian manifold locally has the Euclidean metric structure at the point.

The Schwarzschild manifold is a four-dimensional pseudo-Riemannian manifold. For an arbitrary point on it, say point (ct, $x^r$), r=1,2,3, in the coordinate system {ct, $x^r$}, there must be a locally valid doubly-primed coordinate system {ct", $x'''^r$} in which the point is represented by (0, 0, 0, 0) and the Schwarzschild manifold has the Minkowskian-type line element,

$$ds^2 = c^2 dt''^2 - \delta_{rs} dx''^r dx''^s, \quad r,s=1,2,3, \qquad (7)$$

at the point.

We are familiar with the Minkowskian-type line element. In Eq.(7), ds is a four-dimensional "distance" between two events taking place at neighboring points (0, 0, 0, 0) and (cdt", $dx'''^r$), r=1,2,3. This four-dimensional "distance" is invariant in all coordinate systems. When two events take place at the same spatial point, i.e. $dx''^r$ =0, r=1,2,3, ds signifies the proper time dT by ds=cdT. Setting $dx''^r$ =0 in Eq.(7), we have

$$dT^2 = dt''^2. \qquad (8)$$

When two events take place at the same instant, i.e. dt"=0, ds signifies the proper distance dX by ds=-dX. Setting dt"=0 in Eq.(7), we have

$$dX^2 = \delta_{rs} dx''^r dx''^s. \qquad (9)$$

We shall call dT and dX respectively the gravitational proper time and gravitational proper distance in order to distinguish them from those called the proper time and proper distance in special relativity.

To find the expressions of gravitational proper time dT and gravitational proper distance dX in the coordinate system {ct, $x^r$}, we recall the invariance of $ds^2$ in the doubly-primed coordinate system {ct", $x'''^r$} and the coordinate system {ct, $x^r$},

$$c^2 dt''^2 - \delta_{rs} dx''^r dx''^s = (1 - \frac{r_s}{r}) c^2 dt^2 + g_{rs}(x) dx^r dx^s. \qquad (10)$$

Comparing the terms on the right-hand side and left-hand side of this equation, in consideration of (1) the gravitational proper time is relevant to only dt while the gravitational proper distance to only $dx^r$, r=1,2,3, and (2) the gravitational proper time and gravitational proper distance respectively become dt and $\delta_{rs} dx^r dx^s$ when the spherically symmetric gravitational field is turned off, i.e. M=0, we naturally pick out

$$dt''^2 = (1 - \frac{r_s}{r}) dt^2, \qquad (11)$$

$$\delta_{rs} dx''^r dx''^s = -g_{rs}(x) dx^r dx^s. \qquad (12)$$

Eqs.(8), (9), (11) and (12) imply

$$dT^2 = (1 - \frac{r_s}{r}) dt^2, \qquad (13)$$

$$dX^2 = -g_{rs}(x) dx^r dx^s, \qquad (14)$$

where $g_{rs}(x)$ is in Eq.(6b).

The gravitational proper time and gravitational proper distance are of local concepts. In inertial frame of reference, the gravitational proper time and gravitational proper distance in the spherically symmetric gravitational field or on the Schwarzschild manifold are nothing but, respectively, the time and distance when the spherically symmetric gravitational field is turned off. As being represented in the coordinate system {ct, $x^r$}, they depend only on spatial coordinates $x^r$, r=1,2,3, of the point. The reason for this is that the Schwarzschild manifold is static. In general, the gravitational proper time and gravitational proper distance in a non-static gravitational field will depend on both time and spatial coordinates of the point.

4. The velocity space

When we study the sufficiently small immediate vicinity at the point, (ct, $x^r$), r=1,2,3, in the coordinate system {ct, $x^r$} or (0, 0, 0, 0) in the doubly-primed coordinate system {ct", $x'''^r$}, on the Schwarzschild manifold, or when we study the tangent hyper-plane attached to the point, coordinates ct and $x^r$, r=1,2,3, or coordinates ct"=0 and $x'''^r$=0, r=1,2,3, of the point are to be regarded as constants, while dt and $dx^r$, r=1,2,3, as well as dt" and $dx'''^r$, r=1,2,3, are variables independent of each other.

Awareness of this allows us to find out the velocity space at the point on the Schwarzschild manifold. Dividing Eq.(9) by Eq.(8), we get

$$Y^2 = \delta_{rs} y'''^r y'''^s, \qquad (15)$$

which embodies the velocity space,

$$dY^2 = \delta_{rs} dy'''^r dy'''^s, \quad r,s=1,2,3, \qquad (16)$$

defined in the doubly-primed velocity-coordinate system {$y'''^r$}, r=1,2,3, where $y'''^r = dx'''^r/dt''$ is called the doubly-primed velocity and Y=dX/dT is the gravitational proper speed. Similarly, dividing Eqs.(14) by (13), we have

$$Y^2 = H_{rs}(x) y^r y^s, \quad r,s=1,2,3, \qquad (17)$$

$$H_{rs}(x) = \frac{1}{1 - r_s/r} \begin{pmatrix} 1 + \dfrac{r_s/r}{1 - r_s/r} \dfrac{x^1 x^1}{r^2} & \dfrac{r_s/r}{1 - r_s/r} \dfrac{x^1 x^2}{r^2} & \dfrac{r_s/r}{1 - r_s/r} \dfrac{x^1 x^3}{r^2} \\ \dfrac{r_s/r}{1 - r_s/r} \dfrac{x^2 x^1}{r^2} & 1 + \dfrac{r_s/r}{1 - r_s/r} \dfrac{x^2 x^2}{r^2} & \dfrac{r_s/r}{1 - r_s/r} \dfrac{x^2 x^3}{r^2} \\ \dfrac{r_s/r}{1 - r_s/r} \dfrac{x^3 x^1}{r^2} & \dfrac{r_s/r}{1 - r_s/r} \dfrac{x^3 x^2}{r^2} & 1 + \dfrac{r_s/r}{1 - r_s/r} \dfrac{x^3 x^3}{r^2} \end{pmatrix}, \qquad (18)$$

embodying the velocity space

$$dY^2 = H_{rs}(x) dy^r dy^s, \quad r,s=1,2,3, \qquad (19)$$

in usual velocity-coordinate system {$y^r$}, r=1,2,3, where $y^r = dx^r/dt$ is the Newtonian velocity in the spherically symmetric gravitational field or on the Schwarzschild manifold.

This is the same velocity space represented in two different velocity-coordinate systems. We name this velocity space the velocity space in the spherically symmetric gravitational field or the velocity space on the Schwarzschild manifold. In the velocity space on the Schwarzschild manifold, the doubly-primed and the usual velocity-coordinate systems are connected by

$$dy'''^r = A^r_s(x) dy^s, \quad r,s=1,2,3, \qquad (20)$$

and

$$y'''^r = A^r_s(x) y^s, \quad r,s=1,2,3, \qquad (21)$$

where

$$A^r_s(x)=\beta\{\delta^r_s+ (\beta-1)\frac{x^r x^s}{r^2} \}, \quad r,s=1,2,3, \tag{22}$$

with

$$\beta=\frac{1}{\sqrt{1-r_s/r}} . \tag{23}$$

This is because

$$\delta_{rs}A^r_p(x)A^s_q(x)=H_{pq}(x). \tag{24}$$

When and only when the spherically symmetric gravitational field is turned off, i.e. M=0, the doubly-primed velocity-coordinate system coincides with the usual velocity-coordinate system,

$$A^r_s(x)=\delta^r_s.$$

Since

$$\det[A^r_s(x)]= \beta^4,$$

we have

$$dy''^1 dy''^2 dy''^3 = \beta^4 dy^1 dy^2 dy^3. \tag{25}$$

The following equations come from Eq.(21),

$$y''^1=\beta\{[1+(\beta-1)\frac{x^1 x^1}{r^2}]y^1 + (\beta-1)\frac{x^1 x^2}{r^2} y^2 + (\beta-1)\frac{x^1 x^3}{r^2} y^3\}, \tag{26a}$$

$$y''^2=\beta\{(\beta-1)\frac{x^2 x^1}{r^2} y^1 + [1+(\beta-1)\frac{x^2 x^2}{r^2}]y^2 + (\beta-1)\frac{x^2 x^3}{r^2} y^3\}, \tag{26b}$$

$$y''^3=\beta\{(\beta-1)\frac{x^3 x^1}{r^2} y^1 + (\beta-1)\frac{x^3 x^2}{r^2} y^2 + [1+(\beta-1)\frac{x^3 x^3}{r^2}]y^3\}. \tag{26c}$$

Taking these equations for each of three doubly-primed velocities $y''^r_2$, $y''^r_1$ and $u''^r$, r=1,2,3, and inserting them in the Galilean addition law of the three doubly-primed velocities,

$$y''^r_2 = y''^r_1 - u''^r, \quad r=1,2,3, \tag{27a}$$

we can find

$$y^r_2 = y^r_1 - u^r, \quad r=1,2,3. \tag{27b}$$

Newtonian velocities in the spherically symmetric gravitational field also obey the Galilean velocity addition law. As well as doubly-primed velocities, Newtonian velocities in the spherically symmetric gravitational field range from negative infinity to positive infinity. The velocity space in the spherically symmetric gravitational field is characterized by the Galilean addition law and boundlessness in both the doubly-primed and the usual velocity-coordinate systems.

Eqs.(26a-c) give us

$$(y")^2 = \beta^2 \{(y)^2 + \frac{r_s/r}{1-r_s/r}(\frac{1}{r}\delta_{rs}x^r y^s)^2\}, \tag{28}$$

where

$$(\beta^2-1) = \frac{r_s/r}{1-r_s/r}$$

is employed and $(y")^2 = \delta_{rs} y"^r y"^s$, $(y)^2 = \delta_{rs} y^r y^s$, r,s=1,2,3. Eqs.(25) and (28) will be used below to derive the equilibrium velocity distribution of low-energy particles in the spherically symmetric gravitational field.

5. Equilibrium velocity distribution of low-energy particles

In pre-relativistic mechanics, the velocity space in inertial frame of reference is

$$dY^2 = \delta_{rs} y^r y^s \tag{29}$$

represented in the usual velocity-coordinate system $\{y^r\}$, r=1,2,3, where $y^r = dx^r/dt$ is the well-defined Newtonian velocity in the absence of gravitational field. Relying on two assumptions, (a) the velocity distribution function is spherically symmetric and (b) the x-, y- and z-components of velocity are statistically independent, Maxwell derived his equilibrium velocity distribution [11],

$$P(y^1,y^2,y^3)dy^1 dy^2 dy^3 = N(\frac{m}{2\pi K_B T})^{3/2} \exp[-\frac{m}{2K_B T}(y)^2] dy^1 dy^2 dy^3, \tag{30}$$

where N is the number of particles, m is their mass, T is the temperature, and $K_B$ is determined to be the Boltzmann constant. Assumptions (a) and (b) reflect structural characteristics of the pre-relativistic velocity space in Eq.(29). Maxwell's velocity distribution is non-relativistic. It is only applicable to low-energy particles.

The Euclidean structure of the velocity space on the Schwarzschild manifold, in the doubly-primed velocity-coordinate system, convinces us of the Maxwellian distribution of doubly-primed velocities,

$$P(y"^1,y"^2,y"^3)dy"^1 dy"^2 dy"^3 = N(\frac{m}{2\pi K_B T})^{3/2} \exp[-\frac{m}{2K_B T}(y")^2] dy"^1 dy"^2 dy"^3, \tag{31}$$

for low-energy particles in equilibrium in the spherically symmetric gravitational field. Putting Eqs.(25) and (28) into Eq.(31), we obtain

$$P(y^1,y^2,y^3)dy^1 dy^2 dy^3$$

$$= N\frac{(m/2\pi K_B T)^{3/2}}{(1-r_s/r)^2} \exp\{-\frac{m}{2K_B T(1-r_s/r)}[(y)^2 + \frac{r_s/r}{1-r_s/r}(\frac{1}{r}\delta_{rs}x^r y^s)^2]\} dy^1 dy^2 dy^3, \tag{32}$$

as the equilibrium distribution of Newtonian velocities of low-energy particles located at point $(x^1, x^2, x^3)$ in the spherically symmetric gravitational field.

In our case of the spherically symmetric gravitational field, not loosing generality, we can set $x^3=r$ and $x^1=x^2=0$ for the point on the Schwarzschild manifold and rewrite Eq.(32),

$$P(y^1,y^2,y^3)dy^1 dy^2 dy^3$$

$$=N \frac{(m/2\pi K_B T)^{3/2}}{(1-r_s/r)^2} \exp\{-\frac{m}{2K_B T(1-r_s/r)}[\frac{1}{1-r_s/r}(y^3)^2+(y^1)^2+(y^2)^2]\}dy^1 dy^2 dy^3.$$
(33)

Furthermore, in the cylindrical velocity-coordinate system $\{y^3, y_h, \alpha\}$, $y^1=y_h\cos\alpha$, $y^2=y_h\sin\alpha$, $\alpha\in(0,2\pi)$, we have

$P(y^3, y_h)dy^3 dy_h$

$$=N \frac{(m/2\pi K_B T)^{3/2}}{(1-r_s/r)^2} 2\pi y_h \exp\{-\frac{m}{2K_B T(1-r_s/r)}[\frac{1}{1-r_s/r}(y^3)^2+(y_h)^2]\}dy^3 dy_h,$$
(34)

where $y_h$ is the horizontal component of Newtonian velocity $y^r$, r=1,2,3, in the spherically symmetric gravitational field, $y_h\in(0,\infty)$, $y^3\in(-\infty,\infty)$. The distribution function in Eq.(34) is a function of square of $y^3$, we are then able to write the equilibrium velocity distribution in the form of

$P(y_v, y_h)dy_v dy_h$

$$=N \frac{(m/2\pi K_B T)^{3/2}}{(1-r_s/r)^2} 4\pi y_h \exp\{-\frac{m}{2K_B T(1-r_s/r)}[\frac{1}{1-r_s/r}(y_v)^2+(y_h)^2]\}dy_v dy_h,$$
(35)

where $y_v$ is the vertical component of $y^r$, r=1,2,3, $y_v\in(0,\infty)$.

In the framework of Newton's theory of gravitation and classical statistical mechanics, whether Boltzmann's combinatorial method or Gibbs's canonical distribution method, both lead to the Maxwellian equilibrium velocity distribution for low-energy particles in the spherically symmetric gravitational field. There, the presence of the spherically symmetric gravitational field (actually, of any gravitational field) does not affect the equilibrium velocity distribution of low-energy particles. Here, as we have seen, based on Einstein's theory of gravitation and classical statistical mechanics, the equilibrium velocity distribution of low-energy particles in the spherically symmetric gravitational field is no longer the Maxwellian. The spherical symmetry in the Maxwellian velocity distribution is spoiled by the presence of the spherically symmetric gravitational field. Only when the spherically symmetric gravitational field is turned off, i.e. M=0, the equilibrium velocity distribution of low-energy particles, Eq.(33), reduces to the Maxwellian everywhere. In the spherically symmetric gravitational field produced by mass M, at great distances far from M, $\frac{1}{(1-r_s/r)}\approx 1$, the equilibrium velocity distribution of low-energy particles is approximately the Maxwellian..

6. Leakage of particles in spherically symmetric gravitational field

The obtained equilibrium velocity distribution can be used to calculate the leakage of low-energy particles in the spherically symmetric gravitational field. Suppose we have a small container of volume V placed at point $(x^1, x^2, x^3)=(0, 0, r)$ in the spherically symmetric gravitational field of mass M located at the origin. On the container there is a leak of area $\Delta A$ normal to $x^1$-direction. Low-energy particles in equilibrium inside the container will escape through this leak. We now calculate the rate for these particles to escape.

The number of escaping particles per unit time must be proportional to area $\Delta A$,

$N_{esc}= R_h\Delta A,$ (36)

where $R_h$ is the escape rate per unit time per unit area of the leak. The number of escaping particles per unit time, $N_{esc}$, is also a total number of those particles that hit the leak in a unit time, so the escape rate is

$$R_h = \frac{1}{V} \int_{-\infty}^{\infty} dy^2 \int_{-\infty}^{\infty} dy^3 \int_0^{\infty} y^1 P(y^1, y^2, y^3) dy^1,$$

where $P(y^1, y^2, y^3)$ is in Eq.(33). Completing these integrals, we get

$$R_h = \frac{N}{V} \left(\frac{K_B T}{2\pi m}\right)^{1/2} \left(1 - \frac{r_s}{r}\right)^{1/2}. \tag{37}$$

Instead, if area $\Delta A$ of the leak is normal to $x^3$-direction, we have

$$N_{esc} = R_v \Delta A, \tag{38}$$

and

$$R_v = \frac{1}{V} \int_{-\infty}^{\infty} dy^1 \int_{-\infty}^{\infty} dy^2 \int_0^{\infty} y^3 P(y^1, y^2, y^3) dy^3.$$

From these integrals it arises that

$$R_v = \frac{N}{V} \left(\frac{K_B T}{2\pi m}\right)^{1/2} \left(1 - \frac{r_s}{r}\right). \tag{39}$$

Based on Einstein's theory of gravitation and classical statistical mechanics, in the spherically symmetric gravitational field, the rate for low-energy particles to escape through a leak depends on both position and orientation of the leak. This is interesting. Actually, in the framework of Newton's theory of gravitation and classical statistical mechanics, the escape rate is

$$\frac{N}{V} \left(\frac{K_B T}{2\pi m}\right)^{1/2} \tag{40}$$

depending on neither position nor orientation of the leak. It is worthwhile to mention that the escape rate of low-energy particles near the Schwarzschild radius $r_s$ would be equal to zero no matter how high the temperature or/and density of these particles are.

7. Concluding remarks

In inertial frame of reference, the presence of a spherically symmetric gravitational field makes space and time co-join to be a four-dimensional Schwarzschild manifold. For any point on the Schwarzschild manifold, we have formed the locally valid concepts of gravitational proper time and gravitational proper distance. Relying on these concepts, we have found the velocity space at the point on the Schwarzschild manifold. That velocity space has an affine structure in the usual velocity-coordinate system and the Euclidean structure in the doubly-primed velocity-coordinate system. The transformation between the usual velocity-coordinate system and the doubly-primed velocity-coordinate system has been determined. Recognizing the Maxwellian distribution of doubly-primed velocities and using the determined transformation, we have derived the equilibrium distribution of Newtonian velocities of low-energy particles in the spherically symmetric gravitational field. It is no longer the Maxwellian losing spherical symmetry. We have also used the derived velocity distribution to calculate the rate for low-energy particles in a container with a leak placed in the spherically symmetric gravitational field to escape. The escape rate depends on both position and orientation of the leak. The obtained equilibrium velocity distribution of low-energy particles in the spherically symmetric gravitational field and the calculated escape rate of low-energy particles in the spherically symmetric gravitational field can serve as the tests of Einstein's theory of gravitation, in addition to its other tests, the gravitational red-shift [12-15], the deflection of light in a stellar gravitational field [12,16-17], the perihelion advance of Mercury in solar gravitational field [12,18] and the delay of radar pulses passing near the Sun [19-21]. The tests suggested here can be performed in laboratories of small size on the Earth.

The derived equilibrium velocity distribution of low-energy particles in the spherically symmetric gravitational field is obviously non-relativistic. In fact, the velocity space represented in Eqs.(16) and (19) is non-relativistic. It is only suitable for describing low-energy particles in the spherically symmetric gravitational field. When we study some phenomena connected with equilibrium velocity distribution of high-energy particles in a spherically symmetric gravitational field, such as the rates of nuclear fusion reactions in the Sun and its relevant solar neutrino problem [22-25], we need the relativistic equilibrium velocity distribution in the spherically symmetric gravitational field. For it, we have to consider the relativistic velocity space on the Schwarzschild manifold [26].

Acknowledgment